\documentclass[sigconf]{acmart}

\AtBeginDocument{%
  \providecommand\BibTeX{{%
    \normalfont B\kern-0.5em{\scshape i\kern-0.25em b}\kern-0.8em\TeX}}}

\setcopyright{acmcopyright}
\copyrightyear{2024}
\acmYear{2024}
\acmDOI{x}

\acmConference[Preprint]{}{2024}{}
\usepackage{enumitem}
\usepackage{subfigure}
\usepackage{caption}
\usepackage{stfloats}

\begin{document}

\title{3D-Mirrorcle: Bridging the Virtual and Real through Depth Alignment in AR Mirror Systems}
\author{Yujia Liu}
\authornote{Both authors contributed equally to this research.}
\affiliation{%
  \institution{Tsinghua University}
  \state{Beijing}
  \country{China}}
\email{l-yj22@mails.tsinghua.edu.cn}

\author{Qi Xin}
\authornotemark[1]
\affiliation{%
  \institution{Tsinghua University}
  \state{Beijing}
  \country{China}}
\email{xinq18@mails.tsinghua.edu.cn}

\author{Chenzhuo Xiang}
\affiliation{%
  \institution{Tsinghua University}
  \state{Beijing}
  \country{China}}
\email{xcz21@mails.tsinghua.edu.cn}

\author{Yu Zhang}
\affiliation{%
  \institution{Tsinghua University}
  \state{Beijing}
  \country{China}}
\email{yuzhang-21@mails.tsinghua.edu.cn}

\author{Lun Yiu Nie}
\affiliation{%
  \institution{Tsinghua University}
  \state{Beijing}
  \country{China}}
\email{nlx20@mails.tsinghua.edu.cn}

\author{Yingqing Xu}
\affiliation{%
  \institution{Tsinghua University}
  \state{Beijing}
  \country{China}}
\email{yqxu@tsinghua.edu.cn}

\renewcommand{\shortauthors}{Yujia Liu, et al.}
\renewcommand{\shorttitle}{3D-Mirrorcle}

\begin{abstract}
Smart mirrors have emerged as a new form of augmented reality (AR) interface for home environments. However, due to the parallax in human vision, one major challenge hindering their development is the depth misalignment between the 3D mirror reflection and the 2D screen display. This misalignment causes the display content to appear as if it is floating above the mirror, thereby disrupting the seamless integration of the two components and impacting the overall quality and functionality of the mirror. In this study, we introduce 3D-Mirrorcle, an innovative augmented reality (AR) mirror system that effectively addresses the issue of depth disparity through a hardware-software co-design on a lenticular grating setup. With our implemented real-time position adjustment and depth adaptation algorithms, the screen display can be dynamically aligned to the user's depth perception for a highly realistic and engaging experience. Our method has been validated through a prototype and hands-on user experiments that engaged 36 participants, and the results show significant improvements in terms of accuracy (24.72\% ↑), immersion(31.4\% ↑), and user satisfaction (44.4\% ↑) compared to the existing works.

\end{abstract}

\ccsdesc[500]{Human-centered computing~Visualization systems and tools}
\ccsdesc[500]{Human-centered computing~Mixed / augmented reality}
\ccsdesc[500]{Hardware~Displays and imagers}

\keywords{Augmented Reality, Smart Mirror, Parallax, Depth Alignment}


\begin{teaserfigure}
  \includegraphics[width=\textwidth]{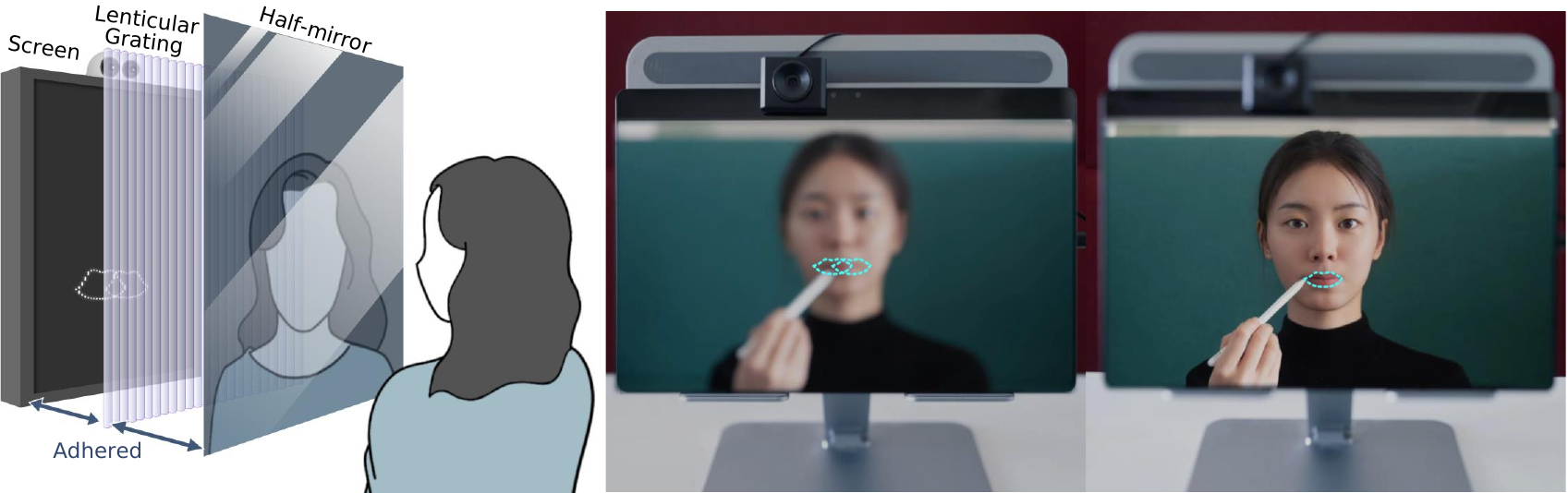}
  \caption {3D-Mirrorcle is an innovative smart mirror system that enables depth-matched AR display on a mirror surface. It uses lenticular gratings, real-time alignment, and segmentation algorithms to present two images on the screen, making each one visible to only one eye. The images merge into a single tracking AR contour, resulting in an immersive experience.}
  \label{fig:1}
\end{teaserfigure}


\maketitle

\section{Introduction}

\begin{figure*}[ht]
    \centering
    \includegraphics[width=0.9\linewidth]{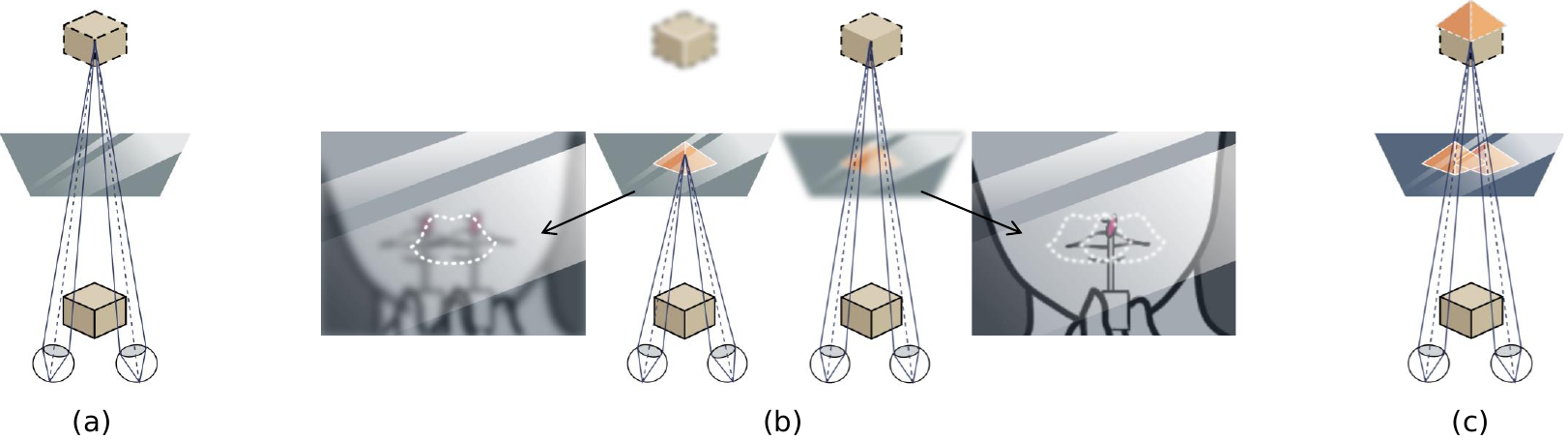}
    \caption{(a) Parallax in human vision. 
    (b) Depth mismatch in 2D smart mirrors: the mirror reflection blurs when viewing the screen display, and the screen blurs vice versa. 
    (c) Two separate images are displayed for each eye to achieve depth perception.}
    \label{fig:why}
\end{figure*}


The recent advancement of Augmented Reality (AR) technology presents immense potential for the creation of novel applications \cite{carmigniani2011augmented, manuri2016survey}. However, the current implementation of AR on smartphones is limited by the restricted screen size and camera capabilities \cite{zhang2020comparative}. Additionally, AR glasses require conspicuous equipment, which could be a hindrance to their widespread adoption \cite{zhuang2019user}. Comparatively, an ideal AR interface should be seamlessly integrated into the real world, without the need for bulky wearables \cite{weiser1999computer,wang2021nanowire,kim2021recent}.

Mirrors, reflecting with depth, not only maintain the authenticity of the real world but also add a virtual dimension, blending reality with fantasy. This unique feature has propelled smart mirrors to prominence as a novel AR interface in home environments. Such mirrors, equipped with augmented displays, find use in various domains, including fitness guidance \cite{anderson2013youmove,hippocrate2017smart} and virtual garment fitting \cite{giovanni2012virtual,gultepe2014real}. To integrate AR into mirror interfaces, two primary methods are employed: the first uses cameras and displays to simulate mirrors, though these often fall short of actual mirrors in resolution, depth perception, and viewing angles \cite{portales2018interacting}; the second method integrates a half-mirror with the display, combining reflective properties and digital content more seamlessly.


However, a "parallax" effect arises due to the slight differences in observation angles between the human eyes when viewing a 3D object \cite{cutting1997eye}, as illustrated in Figure \ref{fig:why} (a). Placing a half-mirror directly on the screen leads to depth mismatch issues, depicted in Figure \ref{fig:why} (b). This configuration fails to align the 2D screen content with the 3D mirror reflection, causing the displayed content to seem floating on the mirror surface. Such a discrepancy severely impairs the user experience.




Consequently, many works have attempted to imbue display content with depth perception. One strategy is to place the screen at a distance from the half-mirror, aligning AR content at the appropriate depth \cite{ushida2002mirror,de2010holocubtile,bimber2005spatial,poston1994virtual,hachet2011toucheo,hilliges2012holodesk}. Another line of approaches add projectors at various system locations in order to achieve depth in AR displays \cite{martinez2014through,butler2011vermeer}. Though these methods succeed in adding depth to AR content, the depth remains fixed and the associated configuration is complicated. This requires users to stay at a constant distance from the setup, which limits their practicality for daily use.


Therefore, our research focus is on developing a compact AR mirror display technology named 3D-Mirrorcle, capable of adjusting depth dynamically. This technology utilizes glasses-free 3D, specifically through a lenticular grating affixed to the screen that allows each eye to see different parts of the image, thereby creating depth perception. As depicted in Figure \ref{fig:why} (c), this configuration causes the brain to perceive an object as behind the screen when the left and right eye views are appropriately offset. Although some studies have explored this concept in mirror AR \cite{lee2012transparent,lee2019optical}, their investigations are primarily theoretical, with a lack of systematic methods and practical implementations.


To bridge this gap, our proposed 3D-Mirrorcle system, as shown in Figure \ref{fig:1}, combines hardware components— a display, a lenticular grating, and a half-mirror — with software algorithms for adaptive mirror reflection alignment and lenticular grating segmentation. Altogether, our hardware-software co-design can generate distinct views for each eye, thereby achieving depth perception. Most importantly, this setup adjusts in real-time to the user's position, enabling a natural and immersive augmented reality experience within the mirror space.



To elevate 3D-Mirrorcle from theory to practice, we evaluate our method through a prototype tested against existing mirror AR displays. User studies involving 36 participants affirmed our system's superior performance in usability, immersion, and acceptability, highlighting 3D-Mirrorcle's potential to integrate AR into daily life.



The key contributions of this paper are:
\vspace{-\topsep}
\vspace{0.3em}
\begin{itemize}
\itemsep0em 
    \item[(1)] A hardware-software co-design framework, featuring a \textit{Mirror Reflection Alignment} algorithm for precise positioning, and a \textit{Lenticular Grating Segmentation} algorithm to achieve accurate depth perception. 
    \item[(2)] A real-world prototype that brings theoretical concepts to practical evaluation, allowing for hands-on testing and improvement.
    \item[(3)] Comprehensive user studies that demonstrate the system's superiority in accuracy, immersion, and user satisfaction compared to existing mirror AR solutions.
\end{itemize}

\section{Related Work}
\subsection{AR in Mirror}
AR interfaces initially used front-facing cameras and screens to blend real and virtual elements. For example, there are AR mirror systems for anatomical education \cite{blum2012mirracle}, motion training \cite{anderson2013youmove}, conductors' practice \cite{salgian2017smart} and stimulate laughter \cite{melder2007affective}.

However, The displays in these systems, unlike real mirrors \cite{portales2018interacting}, suffer from issues like lower resolution, changed perspectives, depth loss, lack of direct eye contact, and delays, making the interaction feel less like viewing one's reflection and more like watching a film.

To maintain the original properties of mirrors, half-mirrors are attached directly to screens, allowing viewers to perceive information displayed on the screen through the mirror. This technology is commonly used in smart mirrors within the Internet of Things (IoT) scenarios displaying date, news, weather, and notifications \cite{alboaneen2020internet, athira2016smart,colantonio2015smart,gomez2017smiwork}. Additionally, there are also mirrors for fitness \cite{hippocrate2017smart, besserer2016fitmirror}, psychological therapy \cite{olowolayemo2018mirror}, and fashion advice \cite{liu2016magic}.

Nevertheless, these 2D smart mirrors cannot seamlessly integrate AR into the mirror space due to the depth mismatch. As a result, people cannot view the mirror and the AR content simultaneously, leading to poor immersion.

To address this problem, some studies focus on enhancing the depth of AR effects. Some systems position half-mirrors at a distance from vertical displays \cite{ushida2002mirror, de2010holocubtile}, and others position the screen and mirror at a specific angle \cite{poston1994virtual, hachet2011toucheo, hilliges2012holodesk}. Additionally, some projects employ projectors and adjust their positions to create depth-enhanced AR displays \cite{martinez2014through,butler2011vermeer}.

To minimize size, some works have employed gratings in displays, such as incorporating a grating between the display and mirror \cite{lee2019optical,lee2020improving} and using gratings on transparent screens \cite{lee2012transparent}. However, such work is rare and mostly at the conceptual stage, lacking systematic methods and applications. We believe that grating is an effective way to address depth mismatch, and it is a mature method in the field of 3D display. Next, we will delve deeper into this display technology from the perspective of the 3D display domain.

\subsection{3D Display}
When viewing, people merge the images obtained by two eyes together, and the brain extracts the binocular disparity as depth perception, which is the biological basis of three-dimensional (3D) display technology \cite{reichelt2010depth}. 3D views are presented by creating differential images for two eyes through various means. 

The commonly used 3D display technologies can be categorized into three types \cite{urey2011state,geng2013three}: stereoscopic direct-view (requiring eyewear), head-mounted and interactive (wearable), and autostereoscopic direct-view (glasses-free). The first two technologies process the images or optical paths directly in front of the eyes through wearable devices, but wearing devices is not only cumbersome but also affects the mirror AR effect. Thus, "glasses-free" is the best option, relying on the structure dividing screen images into multiple columns and alternating between sending them to the left and right eyes, making certain parts of the image hidden to one eye but visible to the other \cite{geng2013three}.

Among the methods for no eyewear 3D systems \cite{ives1902novel,woodgate1997observer,jones1995liquid,toda1995three,cossairt2007occlusion}, refraction-based is the most space efficient, which applies a lenticular grating on the screen to direct light for specific viewing angles, making it visible only from specific angles in front of the screen. Furthermore, there are also techniques for 360-degree 3D image \cite{endo2000cylindrical} and eye position-based generation \cite{huang2019high, omura1998lenticular, isono199250, park1995lenticular, pastoor19973, sexton1999stereoscopic, son2003parameters}. Specifically, the mature 3D display technology of Leia Lume Pad 2 forms the foundation of our system implementation \cite{fattal2021leia}.

\section{Preliminary Study}

\begin{figure*}[h]
    \centering
    \includegraphics[width=\textwidth]{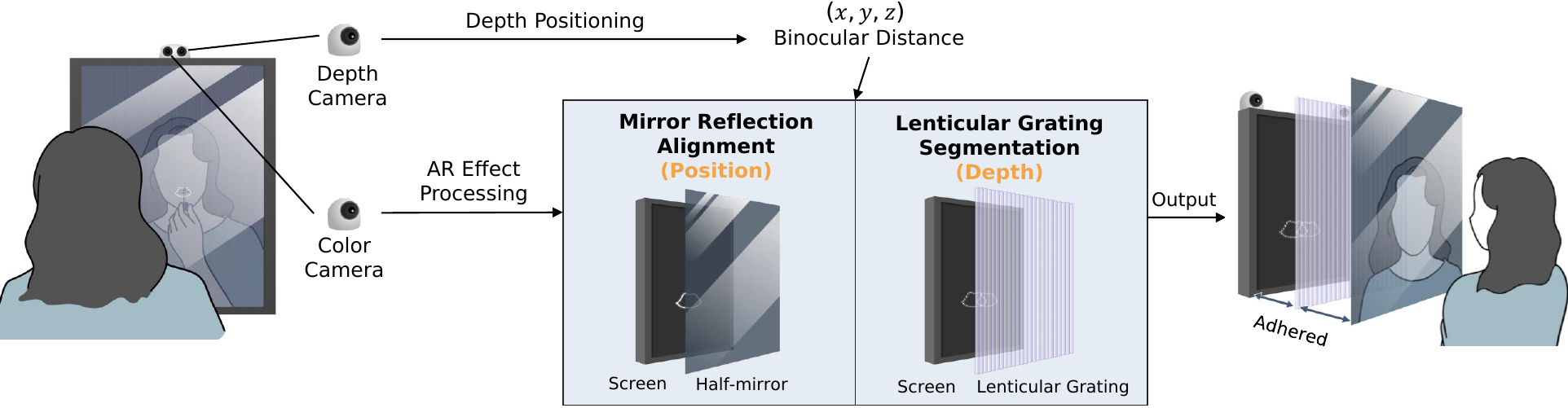}
    \caption{The Pipeline of 3D-Mirrorcle: we locate the coordinates of key points in reality using a depth camera and then process the captured image to achieve an AR effect. After creating a 3D display through lenticular grating and segmented images, viewers can clearly see both their own reflections and the AR effects on the display.}
    \label{fig:pipeline}
\end{figure*}

To comprehend users' habitual interactions with mirrors and establish a set of design principles, we conducted a series of interviews with 12 participants. 

\subsection{Study Design}

In the preliminary study, 12 participants with an average age of 27.5 years ($SD = 5.73$) were involved, including 6 females and 6 males. After gathering the basic information, we conducted semi-structured interviews with each participant, focusing on three research questions.

\begin{itemize}
\item[\textbf{RQ1}:] What are the common scenarios for users' mirror usage?
\item[\textbf{RQ2}:] What are the pros/cons of existing smart mirrors?
\item[\textbf{RQ3}:] What features do users expect from an ideal smart mirror?
\end{itemize}

\subsection{Results}
We utilized a thematic analysis methodology \cite{braun2006using} to analyze the data. Similar codes were combined during the analysis and the results are presented as follows.

\subsubsection{RQ1- Common Scenarios for Users' Daily Mirror Usage}
\label{sec:RQ1}
All of the participants reported using mirrors on a daily basis, but the frequency varied among them. Two participants used mirrors 1-2 times/week, two used mirrors 3-5 times/week, five used mirrors 6-7 times/week, and three used mirrors more than 7 times/week. Females tended to use mirrors more frequently, averaging 6 times/week, while males used mirrors an average of 3 times/week.

The most common reason for using a mirror is to check one's appearance before going out (n=10), where people primarily inspect themselves from a distance of 40-100 cm for about 10 seconds. Additionally, people also use mirrors for makeup or skincare purposes (n=7). While doing so, they prefer adequate lighting conditions and adjust their distance from the mirror as needed. In sporadic situations (n=5), such as experiencing facial discomfort, finding foreign objects on the face, or external reminders, people tend to take quick glances at the mirror or use their mobile camera.

\subsubsection{RQ2- Pros/Cons of Existing Smart Mirrors}


Most of the participants in the study have not used smart mirrors extensively. Among them, 4 participants have never seen smart mirrors, 6 have seen them online, and 2 have seen them in shopping malls. None of the participants own a smart mirror at home.

The overall perception of smart mirrors is negative, with only one participant expressing a willingness to install one at home. The main concerns raised by participants include the lack of usefulness (n=10), privacy issues due to cameras (n=8), high pricing (n=6), and the space and power requirements for installation (n=5).

The prominent points mentioned regarding the advantages include information display (n=3) and whole-home intelligence (n=2). However, due to the current limited functionality of smart mirrors, there are fewer interview results on this aspect.

\label{RQ2}

\subsubsection{RQ3-Expectations for an Ideal Smart Mirror}
In discussions concerning the ideal smart mirror, ``\textit{Immersion}'' emerged as the most commonly cited aspect, with 6 respondents expressing a desire for a "magic mirror" that can offer an immersive experience and transport users to an alternate reality. 4 of them having experience with smart mirrors also noted that the display content should be integrated seamlessly into the mirror's surface. ``\textit{Usefulness}'' was the second most frequently mentioned priority, with 5 respondents noting that current smart mirrors were superfluous. The third priority was the ``\textit{Ease of installation}'', with 3 respondents emphasizing considerations of space usage and wiring, particularly for older users.

\vspace{.3em}
To summarize, according to the preliminary study, people typically use mirrors to check their appearance, especially their faces. However, existing technologies are not meeting user expectations in terms of usefulness, privacy, cost, and ease of installation. To achieve the ideal smart mirror, we can propose the following design principles:

\begin{itemize}
\item[(1)] More immersive displays.
\item[(2)] More simplified structures.
\item[(3)] More diverse applications.
\end{itemize}

\section{3D-Mirrorcle}
Therefore, we introduce 3D-Mirrorcle, a smart mirror system that delivers immersive AR content with a streamlined structure. To achieve consistent depth perception AR effects on mirrors, we propose a hardware-software co-design, as shown in Figure \ref{fig:pipeline}. 

Specifically, 3D-Mirrorcle's hardware design is composed of three layers: screen, lenticular grating, and half-mirror, which are tightly attached without any gaps. Additionally, a depth camera is used to locate the user's eyes, and a color camera is employed to capture images, which are subsequently processed by the AR effects processing unit. For the software implementation, we propose two crucial algorithms: the \textit{Mirror Reflection Alignment} algorithm, which adjusts the image to an appropriate position, and the \textit{Lenticular Grating Segmentation} algorithm, which ensures the image appears at the correct depth. Together, these algorithms enable users to perceive AR content that seamlessly blends with the mirror reflections for an immersive interaction.

\subsection{Parallax}
\begin{figure}[t]
    \centering
    \includegraphics[width=0.69\linewidth]{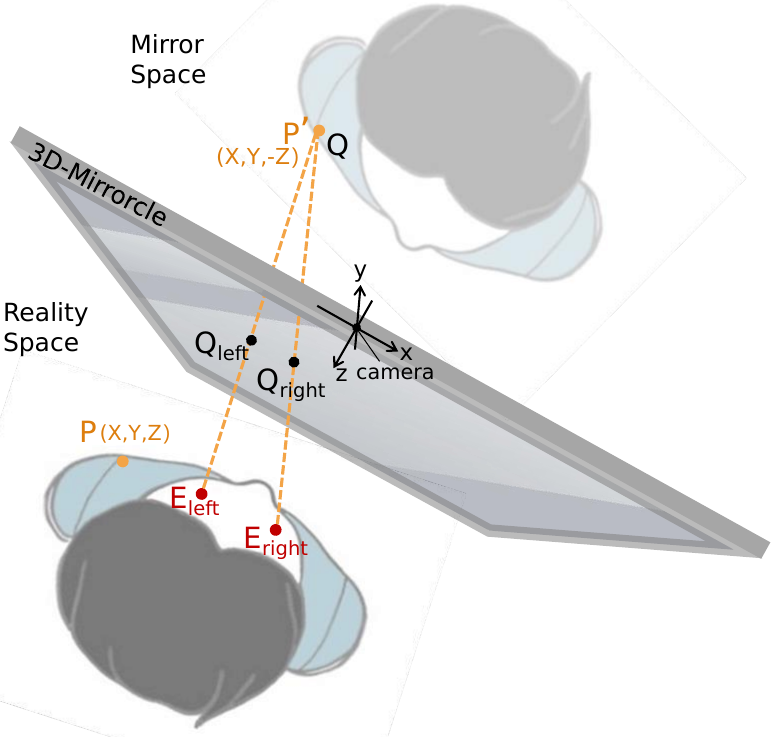}
    \vspace{-7pt}
    \caption{Due to parallax, each eye sees point P' from a distinct angle. To align point Q with P', we create parallax on the screen by showing points $Q_\textit{left}$ and $Q_\textit{right}$ at the sight-screen intersection, each visible to only one eye via lenticular grating, merging to create the illusion of alignment.  }
    \label{fig:AR}
    \vspace{-10pt}
\end{figure}

We begin by giving a formal definition of parallax in human vision. As depicted in Figure \ref{fig:AR}, consider a Cartesian coordinate system where the mirror is situated on the $z=0$ plane, and the position of the depth camera serves as the origin. In this setup, the space with $z<0$ is termed the ``mirror space,'' while the space where actual objects reside (also $z<0$) is referred to as ``real space.''  Let $E_{\textit{left}}=(x_{\textit{left}}, y_{\textit{left}}, z_{\textit{left}})$ denote the position of the user's left eye, and $E_{\textit{right}}=(x_{\textit{right}}, y_{\textit{right}}, z_{\textit{right}})$ the position of the right eye.  Assume that point $P=(X, Y, Z)$ is an arbitrary point in real space, with $P'=(X, Y, -Z)$ representing its mirror reflection. The lines of sight from the left and right eyes to the reflection, denoted as vectors $(E_{\textit{left}}, P')$ and $(E_{\textit{right}}, P')$, are distinct. This difference in viewing angles is what constitutes parallax.

To display a point $Q$ on the screen such that it overlaps with point $P'$ from the viewer's perception, the display must also exhibit parallax. This involves two steps: (1) Divide point $Q$ into two separate points, $Q_{\textit{left}}$ and $Q_{\textit{right}}$, which will be displayed at the intersection points of the viewer’s lines of sight through the screen. (2) Employ a grating to ensure that the left eye can see only $Q_{\textit{left}}$, and the right eye can see only $Q_{\textit{right}}$. To ensure the accurate display of position and depth, we propose a hardware-software co-design methodology to accomplish both tasks effectively.

\subsection{Mirror Reflection Alignment (Position)}
\label{sec:position}
To determine the positions of $Q_{\textit{left}}$ and $Q_{\textit{right}}$ on the screen, we first capture the pixel coordinates of the human eyes using a color camera with a facial keypoint localization algorithm. These coordinates are transformed into spatial coordinates with the depth camera's parameters. Lines extended from the eyes to the mirror image point $P'(X, Y, -Z)$, representing real point $P(X, Y, Z)$, intersect the screen at $Q_{\textit{left}}$ and $Q_{\textit{right}}$, as shown in Figure \ref{fig:AR}. This ensures that point $Q$, as seen by the observer, overlaps with point $P$.

Connect $E_\textit{left} (x_{\textit{left}}, y_{\textit{left}}, z_{\textit{left}})$ to $P' (X, Y, Z)$, setting the spatial line's parameter as $t_{\textit{left}}$. Then, the parametric equation is:

\begin{equation*}
    \left\{
    \begin{aligned}
    x &= x_{\textit{left}} + t_{\textit{left}}(X - x_{\textit{left}}), \\
    y &= y_{\textit{left}} + t_{\textit{left}}(Y - y_{\textit{left}}), \\
    z &= z_{\textit{left}} + t_{\textit{left}}(Z - z_{\textit{left}}),
    \end{aligned}
\right.
\end{equation*}

as z=0, we have \[
t_{\textit{left}} = \frac{-z_{\textit{left}}}{Z - z_{\textit{left}}}.
\]

Therefore, we can compute the coordinates of point \( Q_\textit{left} \):

\begin{equation*}
    Q_\textit{left} = \left( x_\textit{left} + t_\textit{left} \cdot (X - x_\textit{left}),\ y_\textit{left} + t_\textit{left} \cdot (Y - y_\textit{left}),\ 0 \right).
\end{equation*}


The coordinates of \( Q_{\textit{right}} \) can also be computed in the same way, establishing the spatial coordinates for both \( Q_{\textit{left}} \) and \( Q_{\textit{right}} \) with the depth camera as the origin. Given the depth camera's position at the screen's top, offset by \( dis_{camera} \) from the top-left corner, we transform the coordinates to the screen's system as \( Q_{\textit{left}} + (dis_{camera}, 0, 0) \). After applying the resolution conversion factor \( \theta \), the screen position of \( Q_{\textit{left}} \) is computed as \[
(Q_\textit{left} + (dis_{camera}, 0, 0) ) \cdot \theta,
\]
and similarly for \( Q_{\textit{right}} \).

This process provides a method to map each point on the mirror surface to its corresponding location on the screen, facilitating the real-time alignment of AR content to the mirror reflection.

\subsection{Lenticular Grating Segmentation (Depth)}
\label{sec:3D}

To ensure that each eye sees only its intended image, we use a lenticular grating on the screen, which creates interlaced display strips. This arrangement allows each eye to view only half of the screen's pixels. The lenticular grating, composed of parallel columns with uniform diameter, spacing, and height, causes light to diffract. The width of the strips viewed by each eye is determined by the grating’s imaging law, leading to the following relationship\cite{park1995lenticular, wang2012combined}:

\begin{equation}
  f = \frac{r}{n-1}, \quad w = \frac{ef}{l - f},
  \label{equ:1}
\end{equation}
where $f$ is the focal distance of the lenticular grating, $r$ is the radius of curvature of the grating, $n$ is the refractive index of the grating, $e$ is the distance of human eyes, $l$ is the viewing distance, and $w$ is the width of strips displayed on the screen.
Therefore, the left equation of (\ref{equ:1}) indicates that the focal length of the lenticular grating is determined by its curvature and the material's refractive index, affecting how light is focused to create distinct images for each eye. Subsequently, the right equation calculates the width of display strips visible to each eye, which depends on the eye's distance, the focal length of the grating, and the viewing distance, ensuring the correct parallax and depth perception.

\begin{figure}[h]
    \centering
    \includegraphics[width=0.75\linewidth]{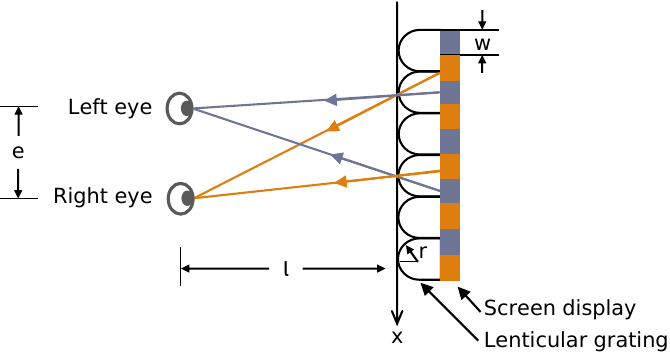}
    \caption{The display is divided into interlaced strips. Through the refraction of a lenticular grating, the left eye can only see the blue strips, while the right eye can only see the orange strips.}
    \label{fig:lenticular}
\end{figure}

Using the lenticular grating imaging formula, we can determine the central positions of the display stripes for the left and right eyes as the viewer's face moves:
\begin{equation}
\label{equation:lenticular}
\begin{aligned}
  x_{\textit{l}}(m) = \frac{x_\textit{left}+x_\textit{right}}{2} + \frac{f\cdot (mT - \frac{1}{2}e)}{l - f} + mT, \\
  \quad x_{\textit{r}}(m) = \frac{x_\textit{left}+x_\textit{right}}{2} + \frac{f\cdot (mT + \frac{1}{2}e)}{l - f} + mT,
\end{aligned}
\end{equation}

where $m$ represents the stripe index to be determined, $T$ is the grating period, $x_\textit{left}$ and $x_\textit{right}$ are aforementioned human eye coordinates. As shown in equation (\ref{equation:lenticular}), when the hardware conditions are fixed, the position of the stripes is determined by the position of the human eyes and the distance $l$ from the screen. Therefore, as the integer $m$ traverses the entire screen, we can obtain the central position of each stripe.

For point $P'$, employing equation (\ref{equation:lenticular}) and the pre-grating refraction coordinates of $Q_{\textit{left}}$ and $Q_{\textit{right}}$ in Section \ref{sec:position}, we determine the specific $m$ value, pinpointing the stripes displaying $Q_{\textit{left}}$ and $Q_{\textit{right}}$. This calculation ensures that the left and right eyes perceive the intended parallax on the screen, aligning with point $P'$.


Note that when the viewer is positioned too close to the screen, the stripe width may exceed the inter-stripe spacing, leading to undesirable overlap of images meant for the left and right eyes. The images are displayed in alternating stripes on the screen. To ensure clear separation, the spacing between successive stripes for the same eye is given by:
\[
x_{\textit{l}}(m) - x_{\textit{l}}(m-1) = \frac{Tl}{l - f},
\]
where \( T \) denotes the grating period, \( l \) the viewing distance, and \( f \) the focal length of the lenticular grating. To prevent image crosstalk, where images for the left and right eyes merge, the spacing should satisfy:
\[
\frac{Tl}{l - f} \geq \frac{2ef}{l - f}.
\]
Thus, the minimum viewing distance \( l \) to avoid crosstalk is:
\[
l \geq \frac{2ef}{T}.
\]
Practically, to prevent displaying overly wide images that cause crosstalk, the pixel width on the screen is rounded down.

\begin{figure*}[ht]
    \centering  
    \subfigure[]{   
        \begin{minipage}{7.7cm}
        \centering    
        \includegraphics[width=\linewidth]{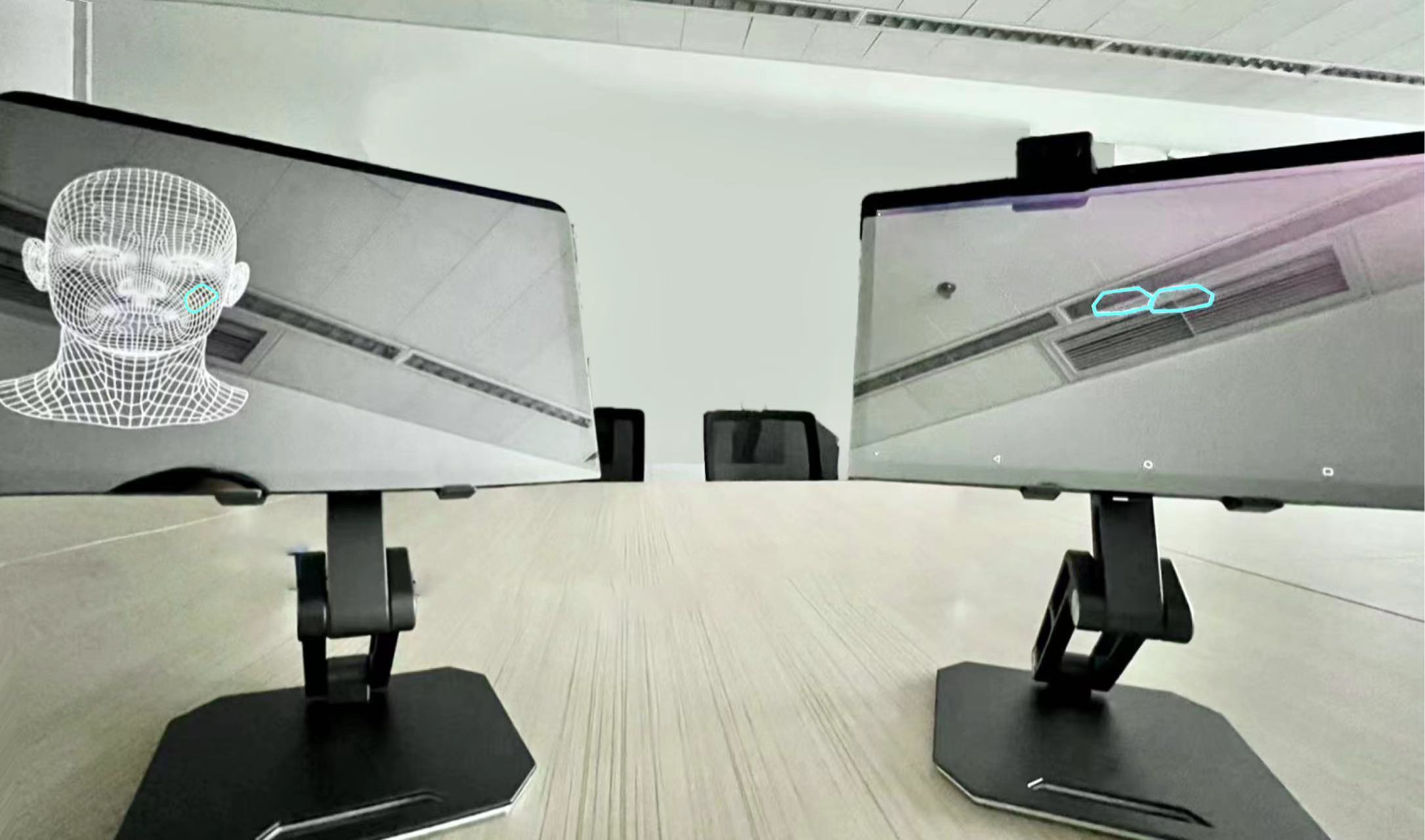}  
        \end{minipage}
    }
    \subfigure[]{   
        \begin{minipage}{4.5cm}
        \centering    
        \includegraphics[width=\linewidth]{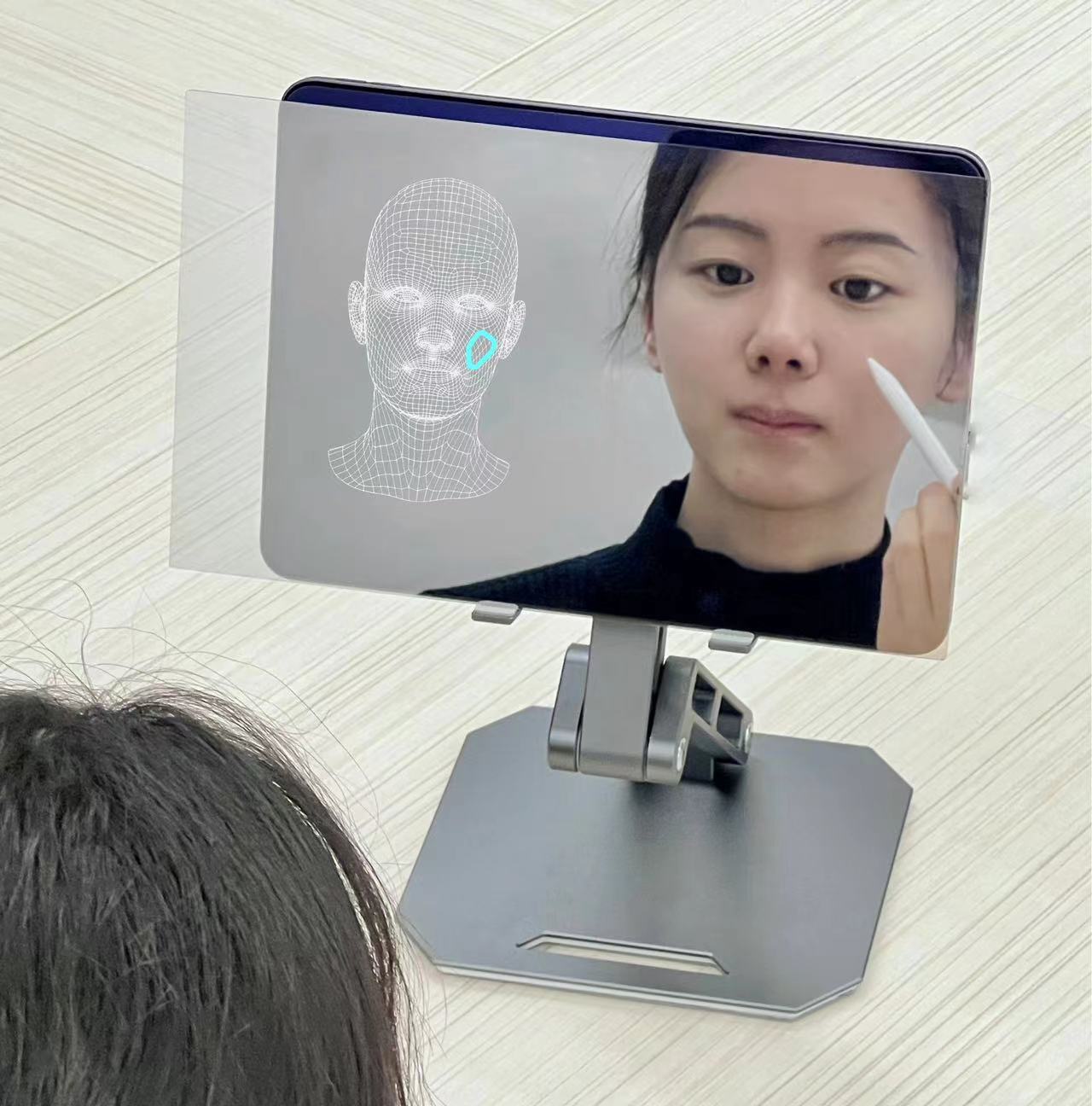}  
        \end{minipage}
    }
    \subfigure[]{ 
        \begin{minipage}{4.5cm}
        \centering    
        \includegraphics[width=\linewidth]{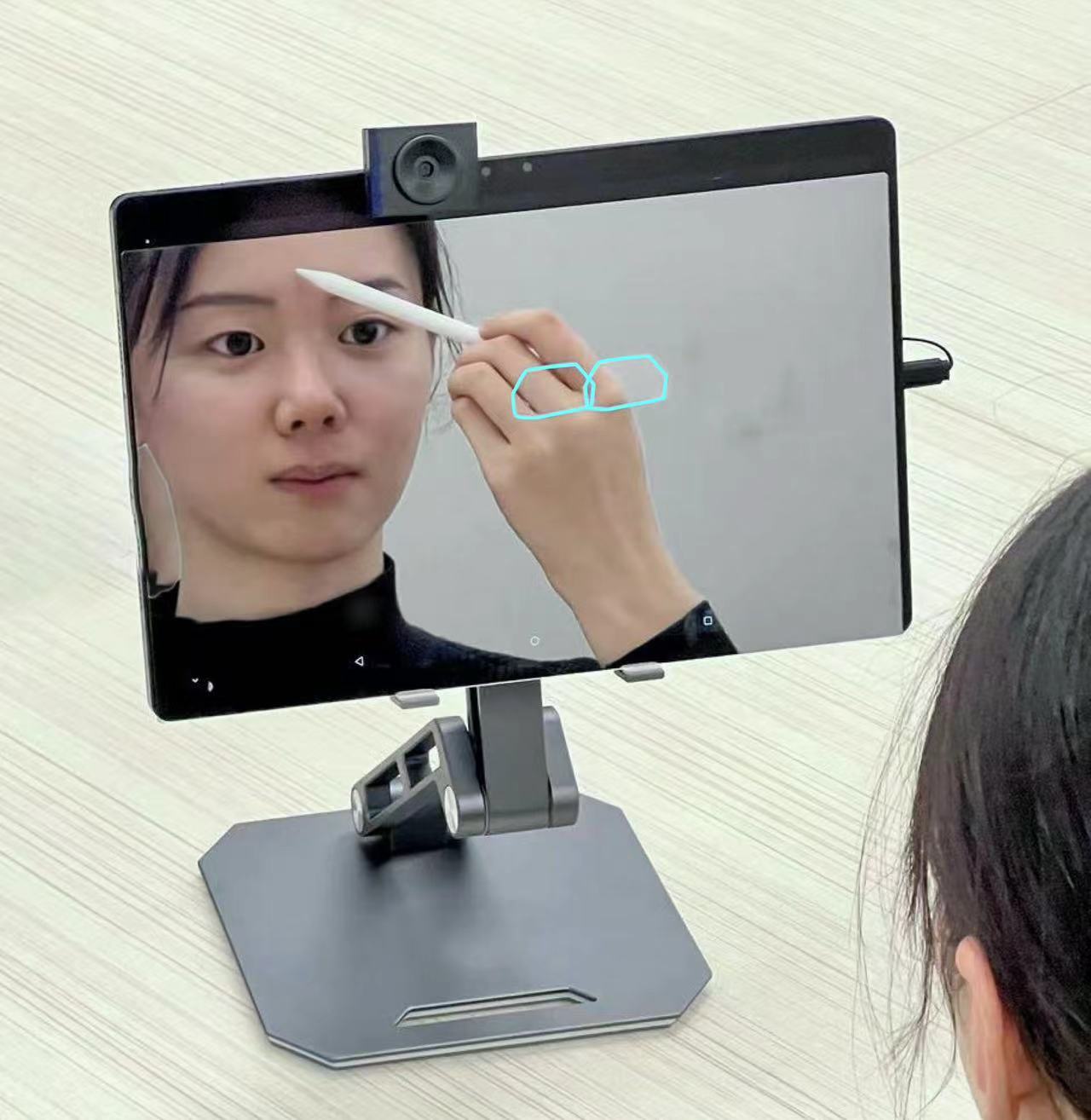}
        \end{minipage}
    }
    \caption{User Study Setups: (a) two mirrors surround the user to compare. (b) left (2D smart mirror): the user can see a guidance image on the screen through a half-mirror. (c) right (3D-Mirrorcle): user's each eye sees one of the images and merges into one depth-aligned image on the face.}    
    \label{fig:AB}    
\end{figure*}

\subsection{Prototype}
To achieve a more effective and lightweight technical validation, we chose the Leia Lume Pad 2\cite{fattal2021leia} as the display hardware. The Lume Pad 2 is equipped with a lenticular sheet for 3D display, enabling high-precision stereoscopic visual effects. In the CNSDK provided by Leia Inc\cite{leiainc_cnsdk}, it is possible to directly render 3D images for specified depths. The Lume Pad 2 also has a front-facing depth camera and a front-facing color camera, facilitating direct binocular positioning and tracking.

\begin{figure}[H]
    \centering
    \includegraphics[width=0.6\linewidth]{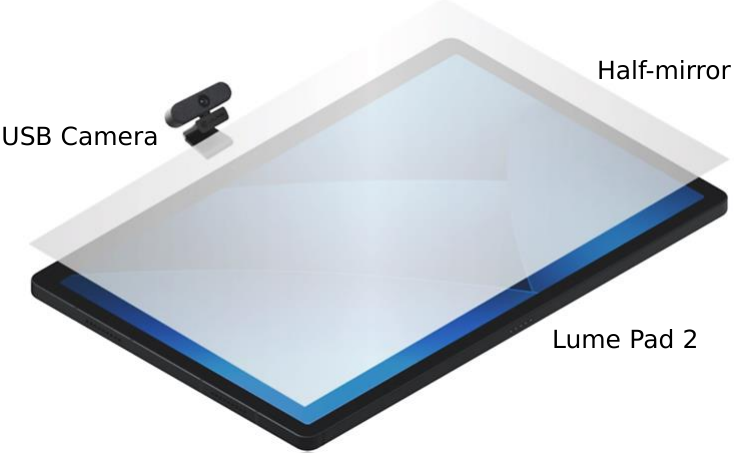}
    \caption{The prototype of 3D-Mirrorcle consists of a Lume Pad 2, a half-mirror, and a USB camera.}
    \label{fig:demo}
\end{figure}

We developed a prototype based on the aforementioned functionality of the Leia Inc CNSDK\cite{leiainc_cnsdk}. By utilizing the built-in depth camera of the Lume Pad 2 for depth positioning and employing the algorithm of MediaPipe\cite{lugaresi2019mediapipe} for keypoints recognition and localization, we implemented the mirror AR algorithm described above. Since the system itself occupies the front-facing color camera, we equipped the Lume Pad 2 with an external USB camera. The 3D-Mirrorcle hardware setup was then completed by simply attaching a half-mirror, as shown in Figure \ref{fig:demo}.

The software is developed as an Android application on the Lume Pad 2 and will be open-source. The Lume Pad 2 features a Qualcomm Snapdragon 888 (2.8GHz) processor, an Adreno 660 GPU, and 14GB of memory.


\section{Evaluation}

One of our main purposes of 3D-Mirrorcle is to enable smart mirrors to interact with the real world, especially the human body. To assess the performance and practicality of our technology, we conducted user experiments with the task of "guiding to delineate contours on faces." Participants completed the task using two display technologies: (a) 2D smart mirror and (b) 3D-Mirrorcle and evaluated their experience based on NASA TLX \cite{hart1988development}.

\subsection{Participants}
We recruited 36 participants (14 males, 22 females), 20-36 years old ($\mu = 27.32$, $\sigma = 4.6$, $\mu$ stands for mean, $\sigma$ stands for standard deviation) \cite{macefield2009specify}. Among them, 30 participants had no prior experience with smart mirrors, while 6 had interacted in shopping malls, with none owning a smart mirror at home. 

\subsection{Study Design}
As Figure \ref{fig:AB} illustrates, on the left is a conventional 2D smart mirror displaying a guidance image on the screen beneath the mirror, serving as the baseline. On the right is the 3D-Mirrorcle, where two images appear on the screen but merge into one at the corresponding depth from the user's perspective.

We instructed participants to outline specific areas on their faces with eyebrow powder, following the guidance provided by each technology. The task involved four areas: the forehead, upper left face, lower right face, and chin. To mitigate the influence of sequence and variations between different areas on the difficulty of outlining, the content displayed on both monitors and the sequence of user experiences were in counterbalanced order. Participants also had a warm-up session to become accustomed to both devices before starting the main task.

\subsection{Measures}
\subsubsection{Objective Evaluation}
We measured two objective indicators: accuracy (\%) and task completion time (s).  
\begin{itemize}
    \item \textit{Accuracy}: The \(IoU\) (intersection-over-union) ratio by comparing the overlap area of the users' drawn contours and the standard area.
    \item \textit{Task completion time}: The duration from the moment the user perceived the display to when the pen left the facial surface.
\end{itemize}

Both these objective indicators were measured by the experimenters without interference in the experimental process.

\subsubsection{Subjective Evaluation}
Our subjective evaluation was based on the NASA TLX questionnaire \cite{hart1988development}, which evaluates Q1: \textit{mental demand}, Q2: \textit{physical demand}, Q3: \textit{temporal demand}, Q4: \textit{performance}, Q5: \textit{effort}, and Q6: \textit{frustration} using a Likert 5-point scale (1=strongly disagree, 5=strongly agree). 
Additionally, to evaluate the virtual environment effectively, we included an assessment of "Q7: immersion" as a key metric \cite{dede2009immersive}. 

\subsection{Results}
\subsubsection{Objective Assessments}
\begin{figure}[h]
    \centering
    \includegraphics[width=0.8\linewidth]{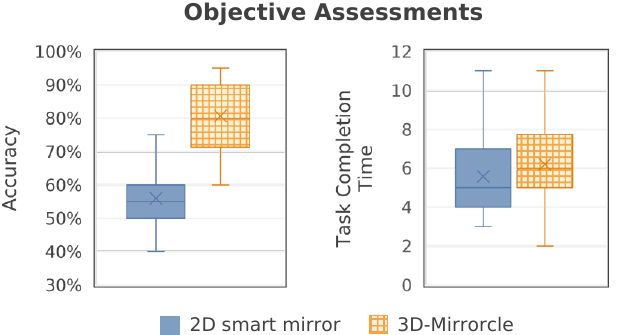}
    \caption{The distribution of objective assessments. Accuracy is calculated by the overlap of the user-drawn and standard area. Task completion time is from the user perceives the display to the pen left the face. Error bars indicate standard error.}
    \label{fig:ob}
\end{figure}

The objective measures of the experiment included accuracy and task completion time, with the results presented in Figure \ref{fig:ob}. The accuracy of the 3D ($\mu = 80.69\%, \sigma = 10.50\%$) was superior to that of the 2D ($\mu = 55.97\%, \sigma = 13.19\%$), and a one-way ANOVA analysis revealed a significant improvement ($F=77.408, p<0.001$, improved 24.72\%). The completion time for the 3D ($\mu = 6.19 seconds, \sigma = 2.227$) was not significantly different from the 2D ($\mu = 5.58 seconds, \sigma = 2.196$) ($F=1.374, p=0.245$).

\subsubsection{Subjective Assessments}
\label{sec:subjective}
\begin{figure}[ht]
    \centering
    \includegraphics[width=\linewidth]{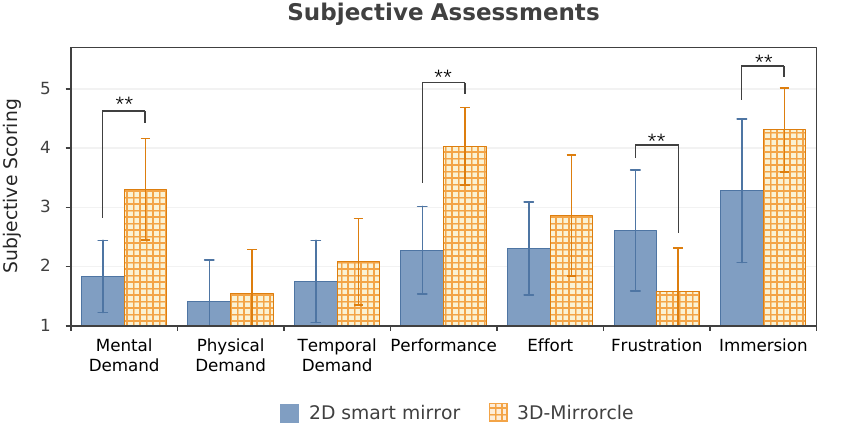}
    \caption{Mean subjective ratings from 1 (strongly disagree) to 5 (strongly agree), measured via NASA TLX and additional metric "immersion." Bridges indicate statistical significance ( *: $p<0.05$, **: $p<0.01$ ). Error bars indicate standard error.}
    \label{fig:sub}
\end{figure}

The subjective ratings of the experiment are shown in Figure \ref{fig:sub}. Compared to the 2D smart mirror, the 3D-Mirrorcle exhibits several significant characteristics: better performance(Q4), less frustration(Q6), better immersion(Q7), but requires a higher mental demand(Q1).

In detail, we compared the scores across each dimension and conducted post-hoc analyses to assess significance. 

(1) For mental demand (Q1), 3D ($\mu=3.31, \sigma=0.86$) was significantly higher than 2D ($\mu=1.83, \sigma=0.61$), with $p<0.001$. 

(2) For physical demand (Q2), 3D ($\mu=1.56, \sigma=0.74$) was slightly lower than 2D ($\mu=1.42, \sigma=0.69$), but not significant ($p=0.41$).

(3) For temporal demand (Q3), 3D ($\mu=2.08, \sigma=0.73$) was slightly higher than 2D ($\mu=1.75, \sigma=0.69$), but not significant ($p=0.051$). 

(4) For performance (Q4), 3D ($\mu=4.03, \sigma=0.65$) was significantly higher than 2D ($\mu=2.28, \sigma=0.74$), with $p<0.001$.

(5) For effort (Q5), 3D ($\mu=2.86, \sigma=1.02$) was slightly higher than 2D ($\mu=2.31, \sigma=0.79$), but not significant ($p=0.012$). 

(6) For frustration (Q6), 3D ($\mu=1.58, \sigma=0.73$) was significantly less than 2D ($\mu=2.61, \sigma=1.02$), with $p<0.001$. 

(7) For immersion (Q7), 3D ($\mu=4.31, \sigma=0.71$) was significantly less than 2D ($\mu=3.28, \sigma=1.21$), with $p<0.001$.

\subsubsection{Overall Satisfaction}
\begin{figure}[ht]
    \centering
    \includegraphics[width=\linewidth]{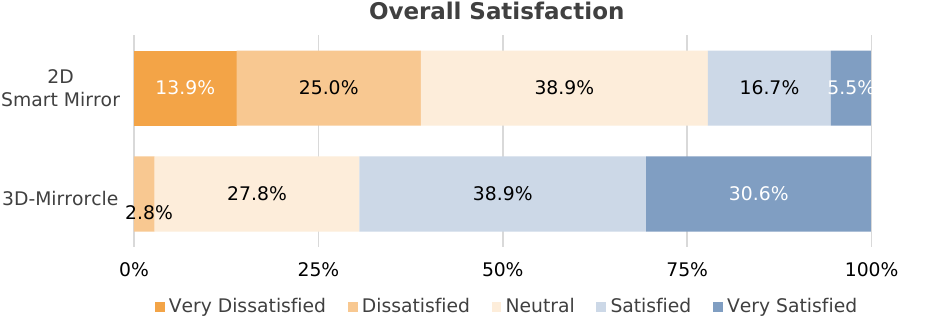}
    \caption{Overall satisfaction ratings from 1 (very dissatisfied) to 5 (very satisfied). 3D-Mirrorcle superior the baseline 44.4\% .}
    \label{fig:satis}
\end{figure}

As illustrated in Figure \ref{fig:satis}, the \textit{overall satisfaction} for 3D ($\mu=3.97,\sigma=0.84$) was significantly higher than for 2D ($\mu=2.75,\sigma=1.08$), with $p<0.001$, improved 44.4\%. 

\begin{figure*}[t]
    \centering
    \includegraphics[width=\textwidth]{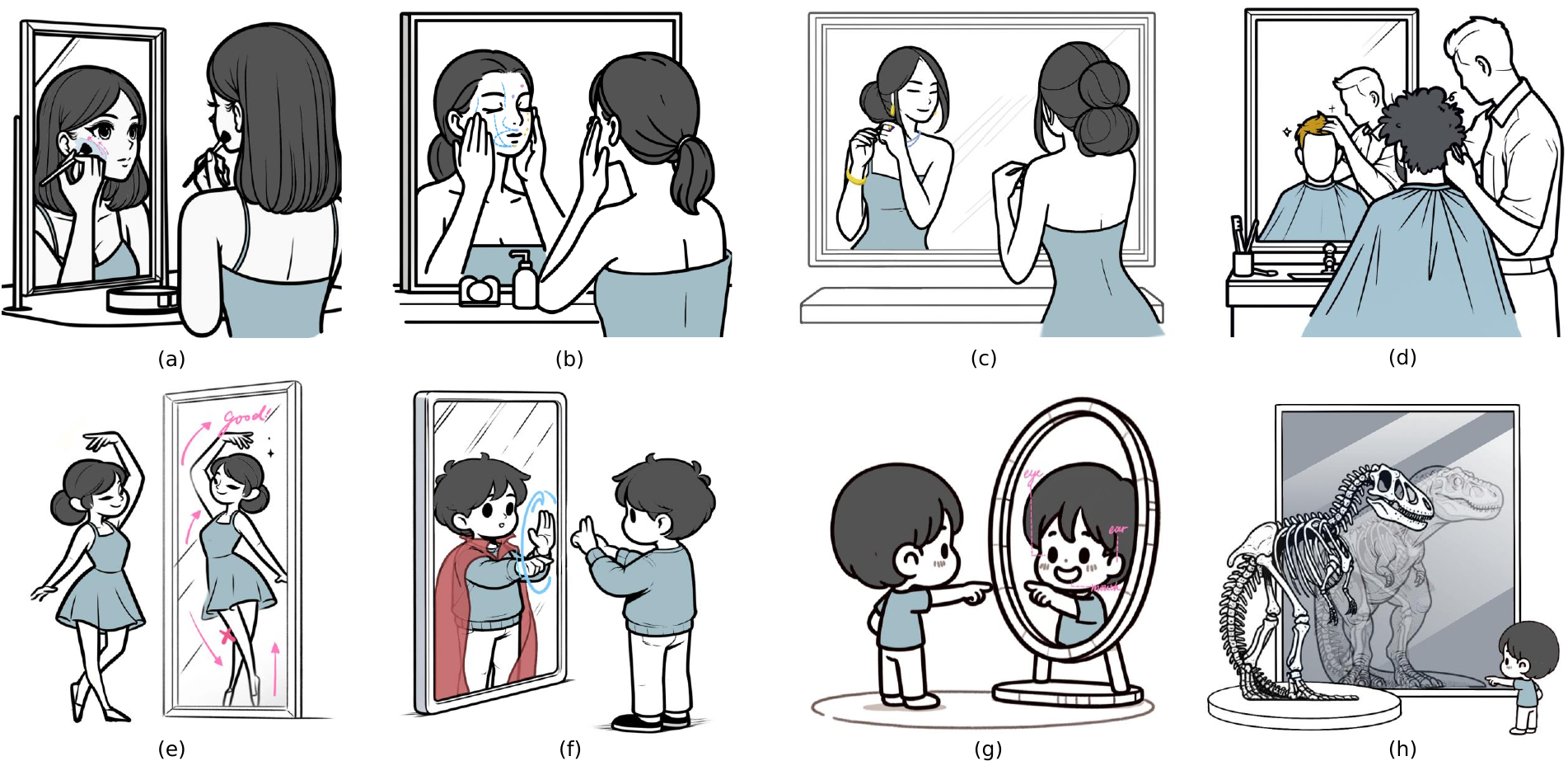}
    \caption{The Potential Applications of 3D-Mirrorcle: (a) Makeup guidance, (b) Facial massage, (c) Jewelry try-on, (d) Hair cutting, (e) Fitness instruction, (f) Interaction game, (g) Embodied education, (h) Exhibition display}
    \label{fig:application}
\end{figure*}

\section{Discussion}
We introduce 3D-Mirrorcle, a smart mirror system that leverages a lenticular grating setup and depth-tracking algorithms, allowing the screen display to be aligned to the user's depth perception for a more realistic interaction experience.

\subsection{Strengths}
Our evaluation shows 3D-Mirrorcle's superiority in both objective and subjective assessments, demonstrating its promise as an innovative AR interface.

\textbf{3D-Mirrorcle is more precise.} 3D-Mirrorcle outperformed the 2D smart mirror baseline in precision by 24.72\%, with users completing tasks relatively quickly. In 2D setups, diagrams could only be shown on the side, leading most users to glance once or twice before relying on their own experience to draw circles on their faces. Users were unsure about the accuracy of their task, yet the process was fast. Statements like \textit{"I don't know if what I'm drawing is right"} (P5, 11, 29, 33) and \textit{"I feel like I'm in a painful drawing class"} (P12) indicate that users experienced poor performance (Q4) and high frustration (Q6). In contrast, the depth-accurate display of 3D-Mirrorcle significantly boosted users' confidence, improving accuracy and also offering quick speeds. Comments like \textit{"I did this very well"} (P2, 5, 10) reflect users' perceptions of high performance (Q4) and low frustration (Q6) with the 3D-Mirrorcle.

\textbf{3D-Mirrorcle is more immersive.} Using a 2D smart mirror requires frequent viewpoint switches, with users commenting, \textit{"I feel like the information is floating on the surface of the screen"} (P2, 19) and \textit{"This is no difference between having a smartphone placed next to the mirror"} (P28, 35). In contrast, the 3D-Mirrorcle displays content with depth, immersing users in a realistic mirror space. Users noted, \textit{"I feel like the information is in the same space as me"} (P5,24) and \textit{"I can complete more complex tasks using this technology"} (P3, 12, 13). Thus, enhanced immersion is a prominent feature of the system, crucial for continuous and smooth task execution.

\textbf{3D-Mirorrcle is more interactive.} Unlike 2D mirrors that display static information, 3D-Mirrorcle integrates with the real environment, offering dynamic 3D content. Users mentioned, \textit{"I think existing AR applications can all be implemented here"} (P7,13) and \textit{"With existing AR, you can't see yourself, but with this technology, you can interact with your reflection in the mirror, which is interesting"} (P24). Therefore, we will discuss the potential applications in detail in the following section.

\subsection{Potential Applications}
\label{sec:application}
The application scenarios dictate how the 3D-Mirrorcle technology integrates into people's daily lives. After user experiments, we also solicited feedback on potential application scenarios.  The summarized applications are depicted in Figure \ref{fig:application}.

\textbf{Makeup Guidance:} 3D-Mirrorcle is ideal for makeup application, where precision and depth perception are crucial. AR technology enhances the experience by providing real-time guidance and visual overlays.

\textbf{Facial Massage:} In facial massage, particularly for locating acupoints in traditional Chinese medicine, 3D-Mirrorcle’s depth-matching AR guides users accurately, enhancing the effectiveness of each session.

\textbf{Jewelry Try-On:} 3D-Mirrorcle allows users to virtually try on pieces of jewelry, viewing them from various angles with realistic depth effects, thereby aiding in better decision-making.

\textbf{Hair Cutting:} For hair styling, 3D-Mirrorcle acts as a communication bridge between customers and stylists, projecting desired hairstyles onto the customer’s reflection for precise adjustments and visualization before the actual cut.

\textbf{Fitness Instruction:} 3D-Mirrorcle transcends traditional fitness mirrors by incorporating interactive, game-like elements into workout routines, making fitness more enjoyable and engaging through dynamic, depth-enhanced instruction.

\textbf{Interactive Gaming:} Leveraging its interactive capabilities, 3D-Mirrorcle supports games that respond to gestures and voice commands, offering immersive experiences like virtual outfit changes, house building, or competing with virtual or real opponents.

\textbf{Embodied Education:} The immersive nature of 3D-Mirrorcle makes it a powerful tool in embodied learning, assisting in activities like identifying body parts, teaching dental hygiene, or interactive language learning.

\textbf{Exhibition Display:} In exhibition settings, 3D-Mirrorcle enriches visitor experiences by extending physical displays into the virtual realm, creating captivating and immersive environments within the mirror space.

\subsection{Limitations and Future Works}
While 3D-Mirrorcle represents a significant advancement in depth-enhanced mirror displays, several limitations have been identified, necessitating future research and development.

\textbf{Limited Number of Users:} The current specifications of the half-mirror and grating limit the viewing experience to single users, precluding simultaneous multi-user interaction. Future research can investigate multi-viewpoint tracking, possibly through the use of spherical gratings, to enable multi-user engagement.

\textbf{Reflection and AR Overlay Limitations:} The fixed reflectance of the half-mirror hinders seamless AR integration, preventing a complete overlay akin to existing AR technologies. Future research into photochromic glass, capable of dynamically altering its transparency, may facilitate smoother transitions between overlay and replacement effects.

\textbf{User Acclimatization:} The unique display approach of 3D-Mirrorcle necessitates user adaptation to simultaneously perceive AR effects and reflections. Future iterations should aim to streamline the hardware design to reduce the adaptation period, thereby making the technology more accessible to a broader audience.

\textbf{Privacy Concerns:} The use of cameras for data capture in 3D-Mirrorcle raises privacy issues. To mitigate these risks, implementing physical camera shutters and enhancing data security with encryption methods is essential for safeguarding user privacy.

\textbf{More Diverse Applications:} With the display system in place, the next step is to expand the range of applications and pinpoint the best use cases for 3D-Mirrorcle, affirming its utility and applicability in everyday situations.

\section{Conclusion}

We presented 3D-Mirrorcle, an innovative system capable of rendering depth-tracking AR effects in a mirror setting. Based on the design principles derived from the pre-study, we carried out a hardware-software co-design, which includes a \textit{Mirror Reflection Alignment} algorithm to ensure accurate positioning, and a \textit{Lenticular Grating Segmentation} algorithm to guarantee precise depth perception. Through user experiments, we demonstrate superior accuracy, enhanced immersion, and satisfaction compared to existing mirror AR technology. This advancement has promising potential in various applications like facial actions, education, games, and exhibitions, which could significantly impact people's daily lives.



\bibliographystyle{ACM-Reference-Format}
\bibliography{ref}

\end{document}